\documentclass[prb,twocolumn,showpacs,floats,eqsecnum,amsmath,amssymb]{revtex4}
\usepackage[dvips]{graphicx}
\begin{document}

\title{Surface Plasmon Polaritons in a Waveguide Composed of
 Weyl Semimetals with Different Symmetries}

\author{S. Oskoui Abdol, B. Abdollahipour\footnote{Corresponding author:
b-abdollahi@tabrizu.ac.ir}, A. Soltani Vala}

\address{Department of condensed matter physics, Faculty of
Physics, University of Tabriz, Tabriz 51666-16471, Iran}
\date{\today}
\begin{abstract}
The peculiar topological properties of Weyl semimetals lead to
unusual and unique optical properties. We investigate novel features
of surface plasmon polaritons in a slot waveguide comprised of two
semi-infinite Weyl semimetals with different symmetries, one with
broken time reversal symmetry and the other one with broken
inversion symmetry. We consider Voigt and Faraday configurations for
surface plasmon polaritons at the interface of the Weyl semimetal
with broken time reversal symmetry. We demonstrate that in the Voigt
configuration this structure supports unidirectional surface plasmon
polariton modes above the bulk plasmon frequency while it shows a
nonreciprocal bidirectional dispersion for surface plasmon
polaritons below the bulk plasmon frequency. In particular, we show
that the chiral nature of the surface plasmon polaritons can be
tuned by the topological properties and chemical potential of the
Weyl semimetals. Moreover, in the Faraday configuration we find a
tunable gap in the surface plasmon polariton dispersion. The studied
structures possessing these exotic features may be employed as one
of the building blocks of the chiral optoelectronic devices.
\end{abstract}
\maketitle
\section{Introduction}

Weyl semimetals (WSMs) have recently attracted intensive attention
due to their exotic band structure \cite{Armitage18}. The anomalous
band structure of WSMs manifests itself in their topological
properties such as protected Weyl nodes and Fermi arc surface states
\cite{Murakami07,Wan11-1}. The conduction and valance bands in WSMs
touch each other at Weyl nodes characterized by linear dispersion
around the Fermi level \cite{Fang03}. The Weyl nodes appear in pair
with different chiralities separated in momentum or energy space in
WSMs with broken time reversal symmetry (TRS) \cite{Burkov11} or
spatial inversion symmetry (SIS) \cite{Halasz12}, respectively. Weyl
semimetal phase has been observed in topologically nontrivial
materials including the nonmagnetic samples such as, TaAs
\cite{Xu15-1}, NbAs \cite{Xu15-2}, NbP \cite{Shekhar15} and magnetic
compounds $Y_2Ir_2O_7$ \cite{Wan11} and $Eu_2Ir_2O_7$
\cite{Sushkov15}. Moreover, the so called type-II Weyl semimetal
phase has been observed recently in materials such as $WTe_2$
\cite{Wu16} and $MoTe_2$ \cite{Jiang17} having open Fermi surfaces.
Some exotic effects such as chiral anomaly \cite{Parameswaran14},
anomalous Hall effect \cite{Xu11,Burkov14} and negative
magnetoresistance \cite{Huang15} arise from the non-trivial
topological band structure of WSMs. In particular, an unusual
optical response emerges in WSMs due to the coupling of the
electrical and magnetical properties originating from the chiral
anomaly
\cite{Zyuzin12,Chen13,Vazifeh13,Ashby13,Ashby14,Halterman18}.

The WSMs are promising materials for photonics and plasmonics
applications due to broad tunability of their chemical potential.
Surface plasmon polaritons (SPPs) at the surface of WSMs have been
studied theoretically
\cite{Zhou15,Zyuzin15,Hofmann16,Kotov18,Tamaya18,Oskoui18} and have
been observed at visible wavelengths in $WTe_2$ \cite{Tan18}. SPPs
are collective electromagnetic and electronic charge excitations
confined to the interface of a conductor with a dielectric
\cite{Maier07}. SPPs are employed in the optoelectronic devices such
as surface plasmon resonance sensor \cite{Homola99} and scanning
near field optical microscopy \cite{Novotny06}. It has been shown
that an unconventional plasmon mode exist in WSMs due to the chiral
anomaly which can be used as a signature of the Weyl semimetal phase
\cite{Zhou15}. The interface of two adjacent WSMs with different
magnetization orientations hosts a low-loss localized guided SPPs
\cite{Zyuzin15}. It has been demonstrated that the SPP dispersion
depends on the Weyl nodes separation in energy or momentum space
\cite{Hofmann16}. In a WSM with broken TRS the Weyl nodes separation
vector acts as an effective external magnetic field. Furthermore, it
has been predicted that a giant nonreciprocal waveguide
electromagnetic modes exist in WSM thin films in the Voigt
configuration \cite{Kotov18}. The thickness of WSM thin film and
dielectric contrast of the outer insulators can be used to
fine-tuning of the SPP dispersion and its nonreciprocal property
\cite{Tamaya18}. Recently, we have studied the SPP dispersion and
its properties in a WSM waveguide comprised of two TRS WSMs in
different configurations \cite{Oskoui18}. In particular, it has been
shown that a tremendous unidirectional SPP modes are hosted by this
structure and can be tuned by the chemical potential and topological
parameters of the WSMs. It is worth to note that these intriguing
features of SPP modes in WSMs emerge without need to application of
high external magnetic fields. On the other hand, SPP modes on the
surface of topological insulators (TIs) \cite{Kane05,Hasan10},
exhibiting a bulk gap and gapless surface states protected by TRS,
as the first member of the family of topological materials have been
investigated theoretically
\cite{Raghu10,Efimkin12,Karch11,Schutky13,Qi14} and experimentally
\cite{Pietro13,Lu19}. Due to identical linear Dirac electronic
dispersion of the surface states in TI and graphene, these materials
exhibit very similar dispersion for SPP modes. But, the
spin-momentum locking in the TIs gives rise to the charge and spin
density waves leading to the spin-coupled surface plasmons or spin
plasmons \cite{Raghu10,Efimkin12}. Moreover, exciting
magneto-optical Kerr effect in TIs by employing a ferromagnetic
coupling or an external magnetic field results in generation of a
novel transverse SPP modes in addition to the usual longitudinal one
\cite{Karch11,Schutky13,Qi14}.

In this paper we study the interplay of the SPP modes exist at the
interfaces of two WSMs with broken TRS and SIS connected via a
dielectric layer in the slot waveguide geometry. We consider both
Voigt and Faraday configurations for SPP propagation at the
interface of the WSM with broken TRS. We demonstrate that interplay
of the SPP modes localized at independent interfaces of two WSMs
with different symmetries leads to unidirectional SPP modes in the
waveguide geometry for the frequencies above the bulk plasmon
frequency. However, we find a nonreciprocal bidirectional dispersion
for SPP modes with frequencies below the bulk plasmon frequency. In
particular, our analysis reveals that these chiral SPP modes can be
tuned by the topological parameters and chemical potentials of two
WSMs. In addition, we observe a gapfull SPP dispersion in the
Faraday configuration. We show that the gap of the dispersion can be
controlled by the physical parameters of the WSMs. These exotic
features originating from intrinsic topological properties of WSMs
may be employed practically in optical devices.

The remainder of the paper is organized as follows: In Section
\ref{S2} we introduce our theoretical model and give necessary
equations for deriving SPP dispersion. Section \ref{S3} is devoted
to presenting results and related discussions. Finally, we end by
giving conclusion in Section \ref{S4}.

\section{Theoretical Model and Equations}\label{S2}

The unique optical response of a WSM originating from the intrinsic
topological nature of it, is characterized by a term $\theta
(\mathbf{r},t) = 2(\mathbf{b}.\mathbf{r} - {b_0}t)$ which is called
\textit{axion~angle} \cite{Vazifeh13}. Here $\mathbf{b}$ expresses
the vector connecting two Weyl points in momentum space of a WSM
with broken TRS, whereas $b_{0}$ is the separation of them in energy
for a WSM with broken SIS. The topological properties of WSMs result
in a modified displacement electric field given by \cite{Hofmann16},
\begin{equation}
D = (\varepsilon _{\infty}  + \frac{4\pi i}{\omega }\sigma
)\mathbf{E} + \frac{{i{e^2}}}{{\pi \hbar \omega }}(\nabla \theta )
\times \mathbf{E} + \frac{{i{e^2}}}{{\pi \hbar c\omega }}\dot \theta
\mathbf{B}\ , \label{eq2.1}
\end{equation}
where $\mathbf{E}$, $\mathbf{B}$, $\varepsilon_{\infty}$ and
$\sigma$ are the electric field, the magnetic field, the static
dielectric constant and the conductivity of WSM, respectively. The
second and third terms of Eq. \ref{eq2.1} originate from the
topological properties of WSM corresponding to the anomalous Hall
effect and the chiral magnetic effect, respectively
\cite{Armitage18}. Whereas, the first term of this equation is the
usual displacement field for a normal metal.

For a WSM with broken TRS (${\textbf{b}}\neq 0$), the chiral anomaly
cause to anisotropic optical response characterized by a dielectric
tensor $\hat{\varepsilon}(\omega )$, with diagonal and off diagonal
terms which we will denote them by ${\varepsilon}(\omega )$ and
${\varepsilon_b}(\omega )$, respectively. Also, a WSM with broken
SIS ($b_{0}\neq0$) represents anisotropic optical response with the
same diagonal terms but having different off diagonal terms
represented by ${\varepsilon_{b_0}}(\omega )$.

The SPP localized at the interface of a WSM with a dielectric is
described by an electric field of the form,
\begin{equation}
\mathbf{E} =
(E_x,E_y,E_z){e^{i{\mathbf{q}}\cdot\mathbf{r_\bot}}}{e^{-\kappa
\left| z \right|}} {e^{ - i\omega t}} ,\label{eq2.2}
\end{equation}
which decays exponentially away from the interface in the $z$
direction and propagates in the interface along the direction of
$\mathbf{q}=(q_x,q_y,0)$. This electric field should be satisfied by
the wave equation,
\begin{equation}
\nabla  \times (\nabla  \times\mathbf{E}) =  -
\frac{1}{{{c^2}}}\frac{{{\partial ^2\mathbf{D}}}}{{\partial {t^2}}}
. \label{eq2.3}
\end{equation}
This equation can be expressed as a matrix equation $\hat{
M}\textbf{E}=0$, with $\hat{ M}$ given by,
\begin{equation}
\hat{M} = \left( {\begin{array}{*{20}{c}}
{q_y^2 - \kappa^2}&{ - {q_x}{q_y}}&{ \mp i{q_x}{\kappa}}\\
{ - {q_x}{q_y}}&{q_x^2 - \kappa^2}&{ \mp i{q_y}{\kappa}}\\
{\mp i{q_x}{\kappa}}&{\mp i{q_y}{\kappa}}&{{q_x}^2 + {q_y}^2}
\end{array}} \right) - \frac{{{\omega ^2}}}{{{c^2}}}\hat
\varepsilon (\omega )\ ,\label{eq2.4}
\end{equation}
with the positive sign for $z<0$ and negative one for $z>0$. The
decay constant ($\kappa$) is determined by the condition
$det(\hat{M})=0$. The decay constant in the WSM side depends on the
relative direction of the vectors $\mathbf{q}$ and $\mathbf{b}$.

\begin{figure}
\centerline{\includegraphics[width=9cm]{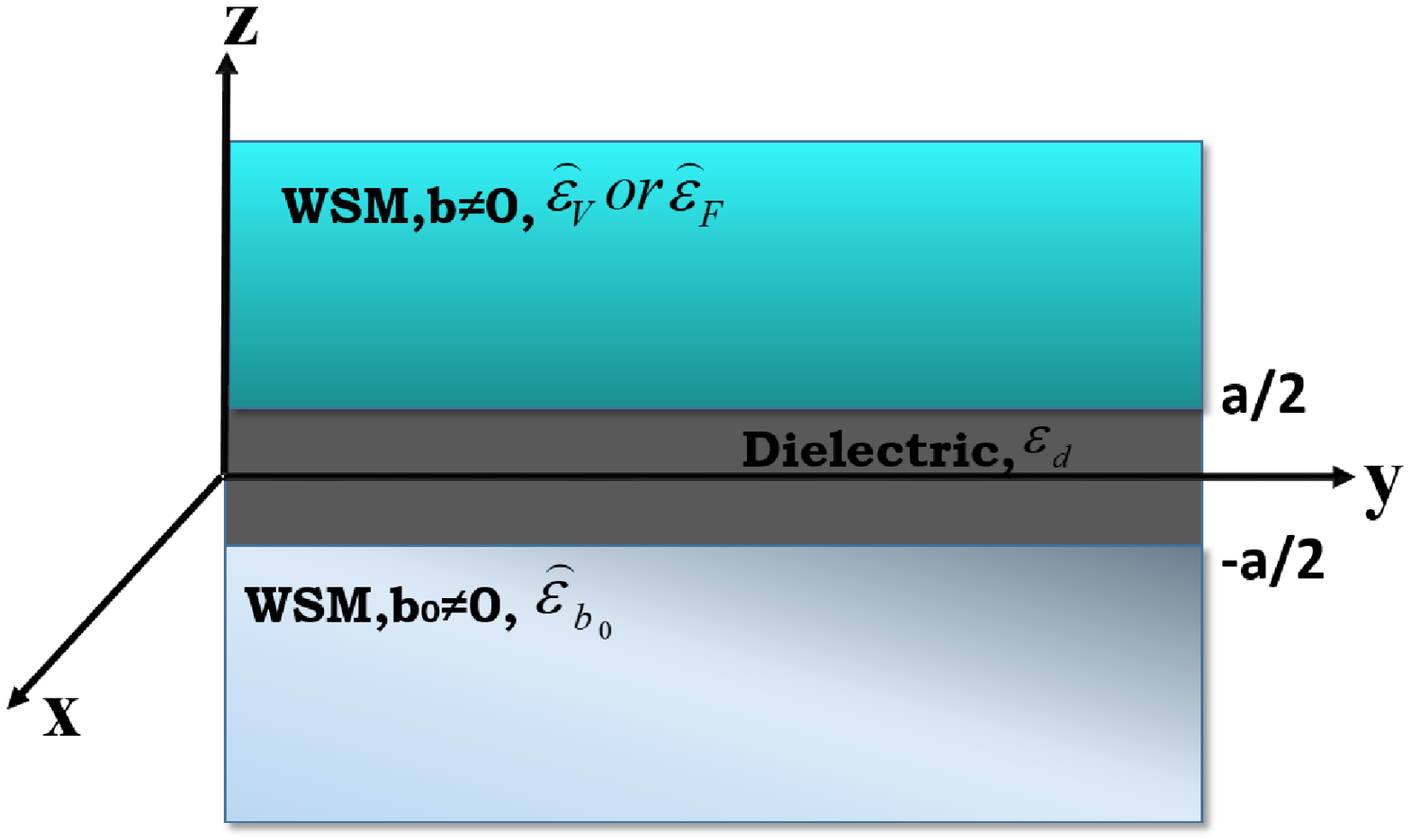}}
\caption{Schematic of the slot waveguide constructed of two
semi-infinite WSMs connected by a dielectric layer with thickness
$a$ and dielectric constant $\varepsilon_{d}$.}\label{fig1}
\end{figure}

The system under consideration is a slot waveguide depicted in Fig.
\ref{fig1}. The slot waveguide has been constructed of two
semi-infinite WSMs, media \textbf{I} and \textbf{III} in the figure,
connected by a dielectric layer (medium \textbf{II}) with thickness
$a$ and a dielectric constant $\varepsilon_{d}$.
%
%

We consider TRS is broken in the upper WSM (medium \textbf{I}) while
in the lower one (medium \textbf{III}) SIS is broken. Relative
direction of the vectors $\mathbf{b}$ and $\mathbf{q}$ results in
different configurations for SPP propagation. We study two different
Voigt and Faraday configurations at the upper interface. In the
Voigt configuration $\mathbf{b}$ is parallel to the surface but
perpendicular to $\mathbf{q}$. By considering $\mathbf{q}=(0,q,0)$
and $\mathbf{b}=(b,0,0 )$ matrix $\hat{M}$ is expressed as,
\begin{equation}
\hat{M_{V}} = \left( {\begin{array}{*{20}{c}}
{q^2 - \kappa _1^2}&0&0\\
0&{- \kappa _1^2}&{ -i{q}{\kappa _1}}\\
0&{-i{q}{\kappa _1}}&{{q}^2}\end{array}} \right) - k_{0}^{2}\hat
\varepsilon_{V} (\omega )\ , \label{eq2.5}
\end{equation}
with
\begin{equation}
\begin{array}{*{20}{c}}
{\hat{\varepsilon}_{V}(\omega ) = \left(
{\begin{array}{*{20}{c}} {\varepsilon}&0&0\\
0&{\varepsilon}&{i\varepsilon _{b}}\\
0&{ -i \varepsilon_{b}}&{\varepsilon}
\end{array}} \right)}\label{eq2.6}
\end{array} .
\end{equation}
In the Faraday configuration $\mathbf{b}=(0,b,0)$ is parallel to the
interface and $\mathbf{q}=(0,q,0)$. In this case the matrix
$\hat{M}$ is given by,
\begin{equation}
\hat{M_{F}} = \left( {\begin{array}{*{20}{c}}
{q^2 - \kappa _1^2}&0&0\\
0&{- \kappa _1^2}&{ -i{q}{\kappa _1}}\\
0&{-i{q}{\kappa _1}}&{{q}^2}\end{array}} \right) - k_{0}^{2} \hat
\varepsilon_{F} (\omega )\ , \label{eq2.7}
\end{equation}
with
\begin{equation}
\hat{\varepsilon}_{F}(\omega )=\left({\begin{array}{*{20}{c}}
\varepsilon&0& i\varepsilon_{b}\\
0&\varepsilon&0\\-i\varepsilon_{b}&0&\varepsilon
\end{array}} \right) .\label{eq2.8}
\end{equation}
Since the lower WSM (medium \textbf{III}) is assumed to be SIS broke
($\mathbf{b}=0, b_0\neq 0$), thus off diagonal terms of the
dielectric tensor results from the third term of Eq. (\ref{eq2.1}).
In this case the matrix $\hat{M}$ can be written in the following
form,
\begin{equation}
\hat{M_{b_0}} = \left( {\begin{array}{*{20}{c}}
{q^2 - \kappa _3^2}&0&0\\
0&{- \kappa _3^2}&{ +i{q}{\kappa _3}}\\
0&{+i{q}{\kappa _3}}&{{q}^2}\end{array}} \right) - k_{0}^{2}\hat
\varepsilon_{b_0} (\omega, q)\ ,\label{eq2.9}
\end{equation}
with
\begin{equation}
\hat{\varepsilon}_{b_0}(\omega, q)=\left({\begin{array}{*{20}{c}}
\varepsilon&-\frac{k_{3}c}{\Omega_{p}}\varepsilon_{b_0}&
+\frac{iq c}{\Omega_{p}}\varepsilon_{b_0}\\
+\frac{k_{3}c}{\Omega_{p}}\varepsilon_{b_0}&\varepsilon&0\\
-\frac{iqc}{\Omega_{p}}\varepsilon_{b_0}&0&\varepsilon
\end{array}} \right) .\label{eq2.10}
\end{equation}
In spite of two former cases, in this case the dielectric tensor has
a direct dependence on the propagation wave vector $\mathbf{q}$. The
diagonal terms of the dielectric tensors in Eqs. \ref{eq2.6},
\ref{eq2.8} and \ref{eq2.10} are equal to ${\varepsilon}(\omega ) =
{\varepsilon _\infty}(1 - \frac{{\Omega _p}^2}{{\omega ^2}})$ and
off diagonal terms are determined by ${\varepsilon_b}(\omega ) =
{\varepsilon _\infty}\frac{\omega _b}{\omega}$ and
${\varepsilon_{b_0}}(\omega ) = {\varepsilon
_\infty}\frac{{\omega_{b_0}}^{2}}{{\omega}^{2}}$. Here ${\Omega_p}^2
=\frac{4\alpha}{3\pi}{(\frac{\mu}{\hbar})^2}$ refers to the bulk
plasmon frequency with $\alpha =\frac{e^2}{\hbar {v_f}{\varepsilon
_\infty}}$, ${\omega _b} = 2{e^2}\left| b \right|/\pi \hbar
{\varepsilon _\infty}$,
${\omega_{b_0}}^2=\frac{2e^2b_{0}\Omega_{p}}{\pi\hbar
c\varepsilon_{\infty}}$ and $k_{0}=\frac{\omega}{c}$ is the wave
vector in the vacuum. It should be noticed that we have ignored the
effect of the carrier scattering in the dielectric tensor of WSMs
due to the very low carrier scattering rates measured in this class
of materials \cite{Shekhar15,Sushkov15}.

Setting the determinant of $\hat{ M}$ matrices in Eqs. \ref{eq2.5},
\ref{eq2.7} and \ref{eq2.9} to zero, the decay constant ${\kappa}$
for aforementioned three different configurations are obtained. For
the Voigt configuration the decay constant should be a positive and
real number and is obtained as,
\begin{equation}
{\begin{array}{*{20}{c}} \kappa_{v+}^{2} = q^2
-{k_{0}^2}\varepsilon, \quad  \kappa_{v-}^{2}= q^2 -
{k_{0}^2}\varepsilon_{v}\ ,
\end{array}}\label{eq2.11}
\end{equation}
where $\varepsilon_{v} = (\varepsilon^2 -
\varepsilon_{b}^2)/\varepsilon$ is the Voigt dielectric function.
For the Faraday configuration it reads,
\begin{equation}
{\begin{array}{*{20}{c}} \kappa_{f\pm}^{2} =  q^2 -
{k_{0}^2}\varepsilon +
k_{0}^2\left(\frac{\varepsilon_{b}^2}{2\varepsilon}\right) \pm
\left[k_{0}^4\frac{\varepsilon_{b}^4}{4\varepsilon^2 }+q^2 k_{0}^2
\frac{\varepsilon_{b}^2}{\varepsilon}\right]^{1/2} \ ,
\end{array}}\label{eq2.12}
\end{equation}
and for the case of $b_0\ne 0$ is given by,
\begin{equation}
{\begin{array}{*{20}{c}} \kappa_{b_0\pm}^{2} =  q^2 -
{k_{0}^2}\varepsilon-\frac{1}{2}k_{0}^{4}\varepsilon_{b_0}^2{(\frac{c}{\Omega_{p}})}^2
\\~~~~~~~~~~~~~~\pm {\frac{1}{2}}[{k_{0}^{8}\varepsilon_{b_0}^4{(\frac{c}{\Omega_{p}})}^4
+4\varepsilon
k_{0}^{6}\varepsilon_{b_0}^{2}{(\frac{c}{\Omega_{p}})}^2}]^{1/2} .
\end{array}}
\label{eq2.13}
\end{equation}
In the dielectric layer which is an isotropic medium the dielectric
tensor is diagonal with elements equal to constant quantity
$\varepsilon_d$. In this medium the decay constant is specified by
$\kappa_{2}=\sqrt{q^{2}-k_{0}^{2}\varepsilon_{d}}$.

{\it SPP modes in the Voigt configuration}: First, we consider the
slot waveguide with Voigt configuration in the upper interface. The
electric field amplitude in three mediums can be written as
combination of the terms in the form of Eq. \ref{eq2.2} with the
appropriate decay constants in the related mediums \cite{Oskoui18}.
Applying the continuity condition for tangential components of the
electric field, denoted by vector $\hat{E}_t$, we arrive at the
following set of linear equations at the upper interface ($z=a/2$),
\begin{widetext}
\begin{equation}
\left({\begin{array}{*{20}{c}}1&0&-e^{\kappa_{2}a/2}&0&-
e^{-\kappa_{2}a/2}&0&0&0\\0&1&0&-e^{\kappa_{2}a/2}&0&-e^{-\kappa_{2}a/2}&0&0\\0&-
\frac{\varepsilon_{1A}}{\kappa_{v-}}\kappa_{2}&0&-
\varepsilon_{0}e^{\kappa_{2}a/2}&0&\varepsilon_{0}e^{-\kappa_{2}a/2}&0&0\\
-\kappa_{v+}&0&-\kappa_{2}e^{\kappa_{2}a/2}&0&\kappa_{2}e^{-\kappa_{2}a/2}&0&0&0
\end{array}} \right)\hat{E}_t=0\ , \label{eq2.14}
\end{equation}
and for the lower interface ($z=-a/2$) it reads,
\begin{equation}
\left({\begin{array}{*{20}{c}}0&0&-
e^{-\kappa_{2}a/2}&0&-e^{\kappa_{2}a/2}&0&1&1\\0&0&0&-
e^{-\kappa_{2}a/2}&0&-e^{\kappa_{2}a/2}&\chi_{1}&\chi_{2}\\
0&0&0&-\varepsilon_{0}k_{0}^{2}e^{-\kappa_{2}a/2}&0&
\varepsilon_{0}k_{0}^{2}e^{\kappa_{2}a/2}&\kappa_{2}
A_{1}&\kappa_{2}A_{2}\\0&0&-\kappa_{2}e^{-\kappa_{2}a/2}&0&\kappa_{2}
e^{\kappa_{2}a/2}&0&\kappa_{b_0+}&\kappa_{b_0-}
\end{array}} \right)\hat{E}_t=0\ ,\label{eq2.15}
\end{equation}
\end{widetext}
where $\varepsilon_{1A}=-\kappa_{v-}(i q \beta_{1}+ \kappa_{v-})$ ,
$A_{1,2}=(i q \theta_{1,2}-\kappa_{b_0\pm} \chi_{1,2})$ with
\small{$\beta_{1}={i(q\kappa_{v-}-k_{0}^{2}\varepsilon_{b})}
/({q^2-k_{0}^{2}\varepsilon})$, $\theta_{1,2}=
{[(-iq\kappa_{b_0\pm})(k_{0}^{2}c\varepsilon_{b_0}\kappa_{b_0\pm}/\Omega_{p})-
(i q
k_{0}^{2}c\varepsilon_{b_0}/\Omega_{p})(-\kappa_{b_0\pm}^2-k_{0}^2\varepsilon)]}/R$},
\small{$\chi_{1,2}={[(iq\kappa_{b_0\pm})(i
qk_{0}^{2}c\varepsilon_{b_0}/\Omega_{p})-(-k_{0}^{2}c\varepsilon_{b_0}\kappa_{b_0\pm/}/\Omega_{p})
(q^2-k_{0}^{2}\varepsilon)]}/R$} and $R={(q^2-k_{0}^2\varepsilon)
(-\kappa_{b_0\pm}^2-k_{0}^2\varepsilon)-(i q \kappa_{b_0\pm})^2}$.

The SPP dispersion relation is obtained by setting the determinant
of the coefficient matrix of Eqs. \ref{eq2.14} and \ref{eq2.15},
which is a $8\times 8$ matrix, to zero. The resultant dispersion
relation is a lengthy equation, not presented here, which we solve
it numerically to obtain SPP dispersion. But, we can find a very
simplified expression for SPP dispersion relation in the nonretarded
limit (long wavelength limit), $q\gg k_{0}$. In this limit all of
the decaying constants tend to the same value of $q$, i.e.
$\kappa_{b_0\pm}=\kappa_{v\pm}=\kappa_{2}=q$. Applying these
simplifications in Eqs. \ref{eq2.14} and \ref{eq2.15}, lead to
$\varepsilon_{1A}=(\varepsilon-\varepsilon_{b})$, $\chi_1=\chi_2=0$
and $A_{1}=A_{2}=0$. Finally, by equating the determinant of the
coefficient matrix of these simplified equations to zero, we can
obtain SPP dispersion relation in the nonretarded limit,
\begin{equation}\varepsilon\left[\varepsilon
-\varepsilon_{b}+\varepsilon_{d} tanh(a q)\right]=0 .\label{eq2.16}
\end{equation}
This relation reveals the chiral nature of the SPP propagation in
the considered structure. Dependence of the dispersion relation to
the sign of the propagation wave vector ($q$) results in different
SPP dispersion for the forward and backward directions. The
asymptotic frequencies for SPP dispersion, frequencies of SPP modes
at the long wavelength limit $|q|\rightarrow\infty$, are obtained by
taking the limit of $a |q|\gg 1$ in Eq. \ref{eq2.16}. This leads to
equations $\varepsilon -\varepsilon_{b}+\varepsilon_{d}=0$ and
$\varepsilon=0$, giving the asymptotic frequencies as,
\begin{equation}
\begin{array}{l}
\omega_{as}^{v\pm}=\frac{\sqrt{\varepsilon_{\infty}^{2}\omega_{b}^{2}+
4\varepsilon_{\infty} \Omega_{p}^{2}(\varepsilon_{d}+
\varepsilon_{\infty})}\pm\varepsilon_{\infty}\omega_{b}}{2(\varepsilon_{d}+
\varepsilon_{\infty})} ,\\
\omega_{as}^{b_0}=\Omega_{p} .\label{eq2.17}
\end{array}
\end{equation}

{\it SPP modes in the Faraday configuration}: Employing the same
procedure explained for Voigt configuration, we can obtain the
following set of linear equations for tangential electric filed
amplitudes at the upper interface ($z=+a/2$),
\begin{widetext}
\begin{equation}
\left({\begin{array}{*{20}{c}}1&1&-e^{\kappa_{2}a/2}&0&-
e^{-\kappa_{2}a/2}&0&0&0\\
-q A_{+}\kappa_{f+}&-q A_{-}\kappa_{f-}&0&-\alpha e^{\kappa_{2}a/2}&0&-
\alpha e^{-\kappa_{2}a/2}&0&0\\
\eta A_{+}&\eta A_{-}&0&-\beta e^{\kappa_{2}a/2}&0&\beta e^{-\kappa_{2}a/2}&0&0\\
-\kappa_{f+}&-\kappa_{f-}&-\kappa_{2}e^{\kappa_{2}a/2}&0&\kappa_{2}e^{-\kappa_{2}a/2}&0&0&0
\end{array}} \right)\hat{E}_{t}=0 \ ,\label{eq2.18}
\end{equation}
and at the lower interface ($z=-a/2$) we have,
\begin{equation}
\left({\begin{array}{*{20}{c}}0&0&-e^{-\kappa_{2}a/2}&0&-
e^{\kappa_{2}a/2}&0&1&1\\
0&0&0&-e^{-\kappa_{2}a/2}&0&-e^{\kappa_{2}a/2}&\chi_{1}&\chi_{2}\\
0&0&0&-\beta e^{-\kappa_{2}a/2}&0&\beta e^{\kappa_{2}a/2}&\kappa_{2}B_{1}&\kappa_{2}B_{2}\\
0&0&-\kappa_{2}e^{-\kappa_{2}a/2}&0&\kappa_{2}e^{\kappa_{2}a/2}&0&\kappa_{b_0+}&\kappa_{b_0-}
\end{array}} \right)\hat{E}_{t}=0 \ ,\label{eq2.19}
\end{equation}
\end{widetext}
where $\beta=k_{0}^{2}\varepsilon_{d}$,
$\alpha=k_{0}^{2}\varepsilon_{b}$,
$A_{\pm}=\frac{q^{2}-\kappa_{f\pm}^{2}-k_{0}^{2}\varepsilon}{\kappa_{f\pm}^{2}+
k_{0}^{2}\varepsilon}$, $\eta=\frac{q
\kappa_{2}\varepsilon}{k_{0}^{2}\varepsilon_{d} \varepsilon_{b}}$,
$B_{1,2}=iq \theta_{1,2}-\kappa_{b_0\pm}\chi_{1,2}$ with
$\theta_{1,2}$ and $\chi_{1,2}$ as defined earlier. Setting the
determinant of the coefficient matrix of Eqs. \ref{eq2.18} and
\ref{eq2.19} to zero, once again gives rise to a lengthy equation
for dispersion relation of the Faraday configuration. In this case,
dispersion relation does not depend on the sign of $q$ and thus the
resultant SPP modes should be reciprocal. In the nonretarded limit,
$q\gg k_{0}$, we have $A_{f\pm}=0$, $B_{1,2}=0$,
$\kappa_{b_0\pm}=\kappa_{f\pm}=\kappa_{2}=q$ and the dispersion
relation reduces to,
\begin{equation}
\varepsilon\left[\varepsilon +\varepsilon_{d} \tanh(a |q|)\right]=0
\ ,\label{eq2.20}
\end{equation}
Again, by taking the long wavelength limit $a|q|\gg 1$, we get the
equations $\varepsilon=0$ and $\varepsilon +\varepsilon_{d}=0$,
which lead to the asymptotic frequencies for Faraday configuration
as,
\begin{equation}
\begin{array}{l}
\omega_{as}^{f}=\Omega_{p} \sqrt{\frac{\varepsilon_{\infty}}
{\varepsilon_{\infty}+\varepsilon_{d}}} \ ,\\
\omega_{as}^{b_0}=\Omega_{p} .
\end{array}\label{eq2.21}
\end{equation}
In the next section we present our numerical results for the SPP
dispersion for both Voigt and Faraday configurations. We will show
that topological properties pave as a feasible way for stable tuning
of SPP propagation without need for application of an external
magnetic field.
%
\begin{figure}
\centerline{\includegraphics[width=9cm]{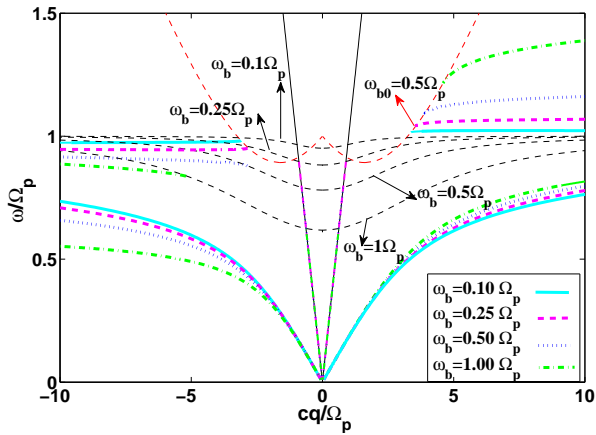}} \caption{SPP
dispersion of the slot waveguide in Voigt configuration at the upper
interface with ${\varepsilon_\infty}=13$, $E_f=0.15~eV$,
$v_f=10^6~m/s$, ${\Omega_p}=60.92~THz$, ${\varepsilon_d}=1.0$ and
for $\omega_{b0}=0.5~\Omega_{p}$, $\omega_b=0.1,~0.25,~0.5,~1.0~
\Omega_p$. The bulk plasmon dispersions related to the upper WSM in
Voigt configuration are indicated by the thin black dash lines and
that for WSM with broken SIS ($b_0\neq0$) is shown by the thin red
dash line.}\label{fig2}
\end{figure}
%
%
\begin{figure*}
\centerline{\includegraphics[height=9cm,width=18cm]{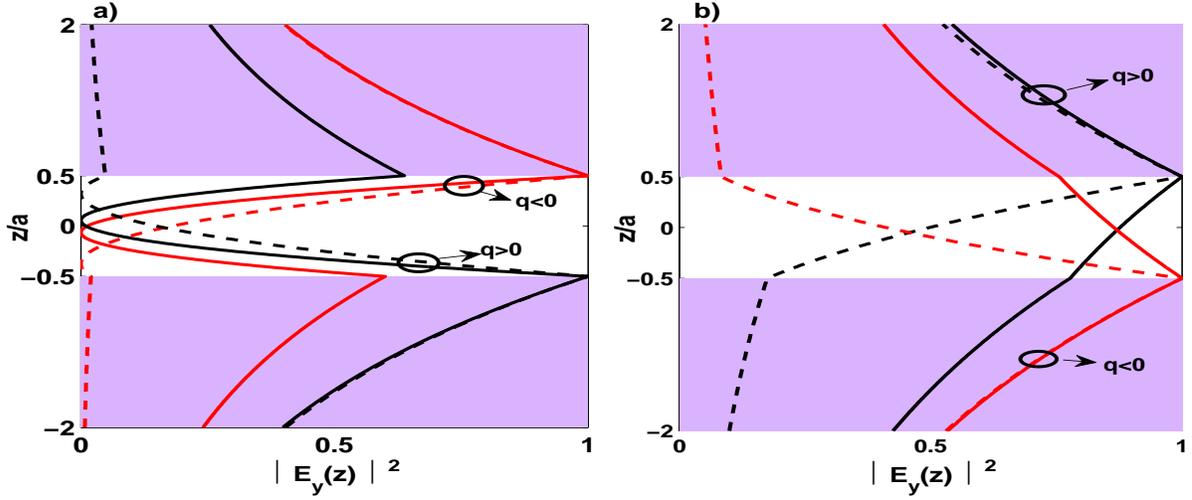}}
\caption{The normalized electric field intensity of SPP modes of a)
lower and b) higher bands as a function of $z$ coordinate for the
waveguide width $a= 0.1~\mu m$ and different
$\omega_{b}=0.1,~1.0~\Omega_{p}$ that have been shown by solid and
dash lines, respectively. The other parameters are same as Fig.
\ref{fig2}. The black lines denote the electric field profile for
positive wave number $(q>0)$ and the red ones for negative wave
number $(q<0)$.}\label{fig3}
\end{figure*}
%
%
%
\begin{figure*}
\centerline{\includegraphics[height=8cm,width=18cm]{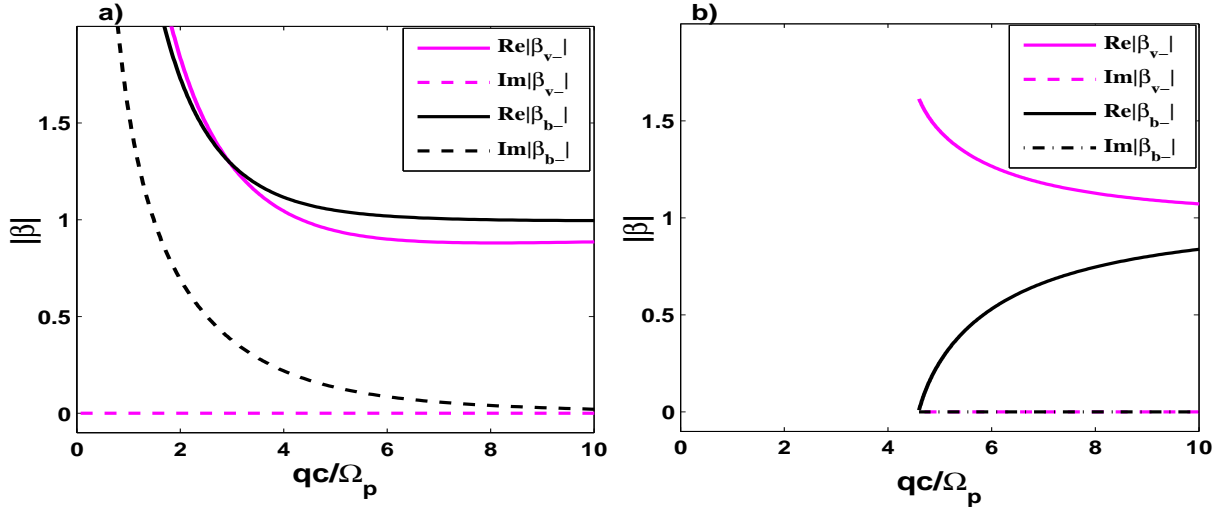}}
\caption{Real and imaginary parts of the reduced decay constants of
SPP modes for a) lower and b) higher bands for $q>0$ as a function
of $q$ shown respectively by solid and dash lines with $a=0.1~\mu
m$, $\omega_{b}=1~\Omega_{p}$ and
$\omega_{b_0}=0.5~\Omega_{p}$.}\label{fig4}
\end{figure*}
%
%
\begin{figure*}
\centerline{\includegraphics[height=9cm,width=18cm]{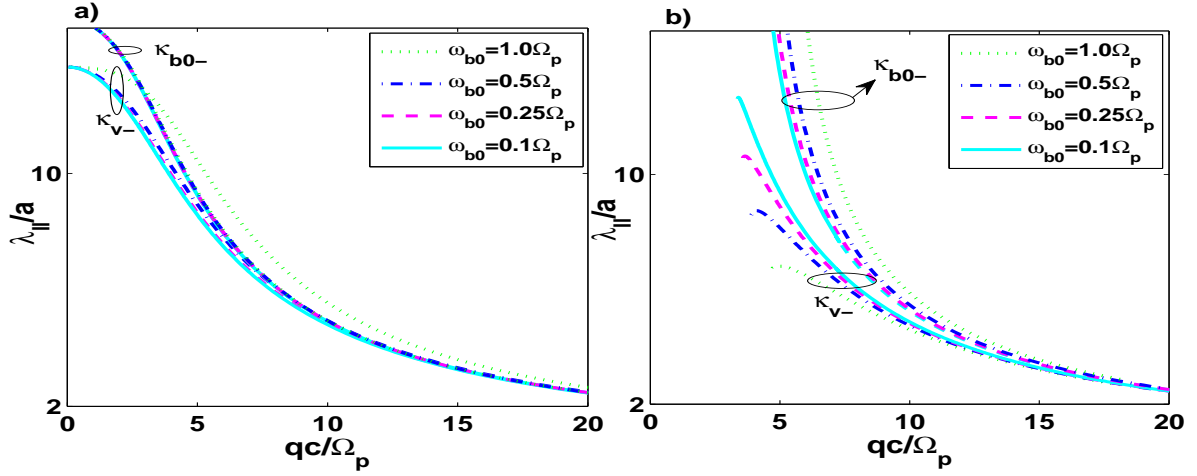}}
\caption{The normalized localization length versus SPP wave vector
for a) lower and b) higher SPP bands with $a=0.1~\mu m$,
$\omega_{b0} = 0.5~\Omega_p$ and for different $\omega_b = 0.1,~
0.25,~0.5,~1~\Omega_p$.}\label{fig5}
\end{figure*}
%
%
\begin{figure}
\centerline{\includegraphics[width=9cm]{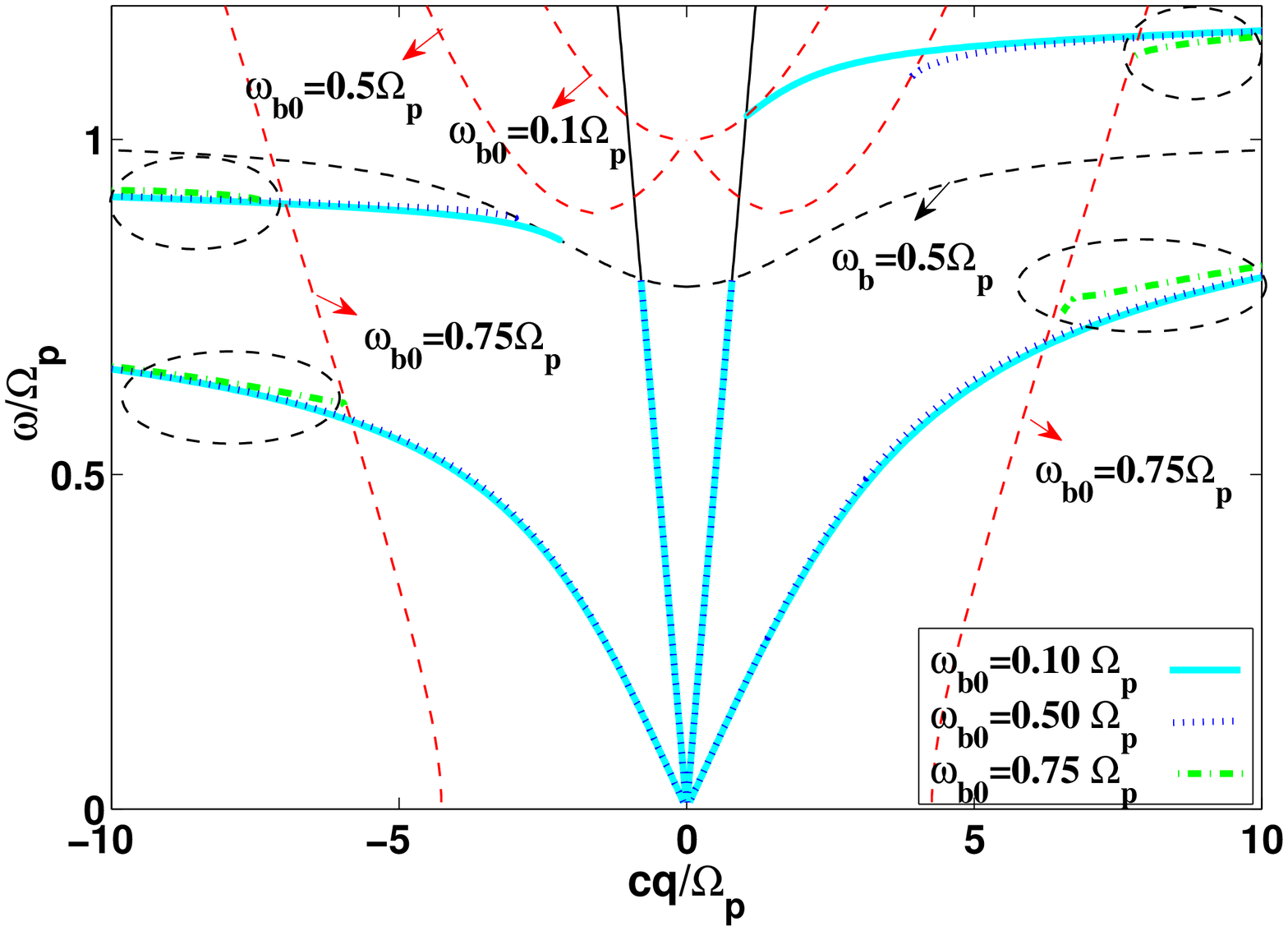}} \caption{SPP
dispersion of the slot waveguide in Voigt configuration at the upper
interface for $\omega_{b}=0.5~\Omega_{p}$ , different
$\omega_{b0}=0.1,~0.5,~0.75~\Omega_p$ and the other parameters same
as Fig. \ref{fig2}. The bulk plasmon dispersions related to the
upper WSM in Voigt configuration and the lower WSM with broken SIS
($b_0\neq0$) are denoted by the thin red dash line and thin black
dash line, respectively.} \label{fig6}
\end{figure}
\section{Results and discussion}\label{S3}
%
Numerical solution of the dispersion relations give dispersion
curves for SPP modes in the slot waveguide  with Voigt and Faraday
configurations. In numerical calculation we adopt the measured
parameters for $Eu_2Ir_2O_7$ as a typical parameters for a WSM:
$\varepsilon_\infty = 13$, $E_f = 0.15~\textit{eV}$ , $v_f =
10^{6}~\textit{m/s}$, $\Omega_p = 60.92~{\textit THz}$, which have
been used in several studies
\cite{Hofmann16,Kotov18,Tamaya18,Oskoui18}.

{\it The slot waveguide with Voigt configuration:} First we consider
the case of Voigt configuration at the upper interface. In Fig.
\ref{fig2} we have exhibited the SPP dispersion in Voigt
configuration for both $q
> 0$ and $q < 0$ for different values of $\omega_b=0.1,~0.25,~0.5,~
1.0~\Omega_{p}$ and $\omega_{b_0}=0.5~\Omega_{p}$. As it is expected
from the dispersion relation (Eq. \ref{eq2.16}) the SPP dispersion
is nonreciprocal in this case and depends on the propagation
direction. There are higher and lower SPP dispersion bands for both
$q>0$ and $q<0$. In addition, there exist a reciprocal SPP
dispersion band lying on the light line corresponding to the lower
WSM with broken SIS. The lower SPP dispersion curves start from the
zero frequency and continuously tend to their asymptotic frequency
given by $\omega_{as}^{v-}$ in Eq. \ref{eq2.17}. The higher SPP
dispersion curves terminate when intersect with the bulk plasmon
modes which have been indicated by the black thin dash lines and red
thin dash line for the upper and lower WSMs, respectively. These
dispersion curves approach to their asymptotic frequencies given by
$\omega_{as}^{v+}$ in Eq. \ref{eq2.17} for $q>0$ and
$\omega_{as}^{b_0}$ for $q<0$. The asymptotic frequencies depend on
the physical parameters of WSMs such as Weyl nodes separation vector
($\mathbf{b}$) and chemical potential ($\mu$). Thus, it is obvious
that we can tune the SPP modes by changing the WSMs parameters. The
dispersion curves lying on the light line start from the zero
frequency and end when coincide with bulk plasmon modes belonging to
the lower WSM. It is remarkable that the SPP modes with frequencies
below the $\Omega_p$ are nonreciprocal but propagate bidirectional
while the SPP modes with frequencies above the bulk plasmon
frequency are chiral and propagate unidirectional. This fascinating
result is in contrast with the SPP modes hosted by a single
interface of a WSM and a dielectric in Voigt configuration
\cite{Hofmann16}. The later structure supports SPP modes for $q<0$
with frequencies above the bulk plasmon frequency, while these modes
disappear in the waveguide structure due to their propagation in the
bulk of the lower WSM with broken SIS. This is a striking result
which shows a profound tunability of SPP modes in the waveguide
structure by involving second WSM with broken SIS. By increasing
$\omega_b$ dispersion curves for $q>0$ shift toward higher
frequencies. This shift is more significant for curves above the
bulk plasmon frequency. For $q<0$ the dispersion curves show an
inverse behavior and their frequencies decrease by increasing
$\omega_b$.

The normalized intensity of $y$ component of the electric field for
lower and higher bands of SPP dispersion have been depicted
respectively in Figs. \ref{fig3} (a) and (b) as a function of $z$
position. The profile of the electric field for $\omega_{b}=0.1~
\Omega_{p}$ (at frequencies $46.42~THz$ for $q>0$ and $44.69~THz$
for $q<0$) and $\omega_{b}=1.0~\Omega_{p}$ (at frequencies
$49.70~THz$ for $q>0$ and $33.62~THz$ for $q<0$) are indicated by
solid and dash lines, respectively. The electric field profile in
Figs. \ref{fig3} (a) and (b) reveal that the SPP modes related to
positive wave number, the black lines, for lower (higher) band is
mostly localized at the lower (higher) interface while the SPP modes
with negative wave number, the red lines, have inverse behavior. A
remarkable feature is that the SPP modes become highly confined to
the corresponding interfaces by increasing $\omega_{b}$.

Decay constants for SPP modes of the bands with frequencies below
the bulk plasmon frequency are pure real and positive in the upper
WSM with Voigt configuration ($\kappa_{v\pm}$) which ensures
decaying of the electric field away from the interface while those
for the lower WSM with broken SIS ($\kappa_{b_0\pm}$) are complex
conjugates of each other which provide the electric field to
oscillatory decay into the lower WSM. Therefore, these SSP modes are
categorized as generalized surface waves \cite{Oskoui18}. On the
other hand, decay constants for SPP modes having frequencies above
the bulk plasmon frequency (only exist for $q>0$) are real and
positive, which these SPP modes are known as normal surface waves.
To illustrate dependence of the decaying constants on wave vector,
we have plotted the real and imaginary parts of the reduced decay
constants $\beta_{v-} = \kappa_{v-}/q$ and $\beta_{b_0-} =
\kappa_{b_0-}/q$ as a function of $q$ for lower and higher bands of
SPP dispersion for $q>0$ in Figs. \ref{fig4} (a) and (b),
respectively. For lower band, Fig. \ref{fig4} (a), $\beta_{v-}$ is
real, while $\beta_{b_0-}$ is complex with an imaginary part
approaching to zero in the limit of large wave vectors. But, both of
these decaying constants are real and positive for SPP modes of the
upper band (Fig. \ref{fig4} (b)) and tend to a same constant value
($\beta=1$) at large wave vectors.

Moreover, the localization length ($\lambda_{ll}=1/Re(\kappa)$)
normalized to the waveguide width ($a$) have been plotted against
wave vector in Fig. \ref{fig5} for higher and lower SPP bands for
different $\omega_{b}$. By decreasing $\omega_{b}$, the localization
length in the upper WSM ($\kappa_{v-}$) for lower (higher) SPP band
decreases (increases), while the localization length in the lower
WSM for lower (higher) SPP band remain intact (decreases) at small
wave vectors. All of $\lambda_{ll}$ for both the higher and lower
bands approach to $\lambda_{ll}=2a$ at large wave vectors.

\begin{figure*}
\centerline{\includegraphics[height=9cm,width=18cm]{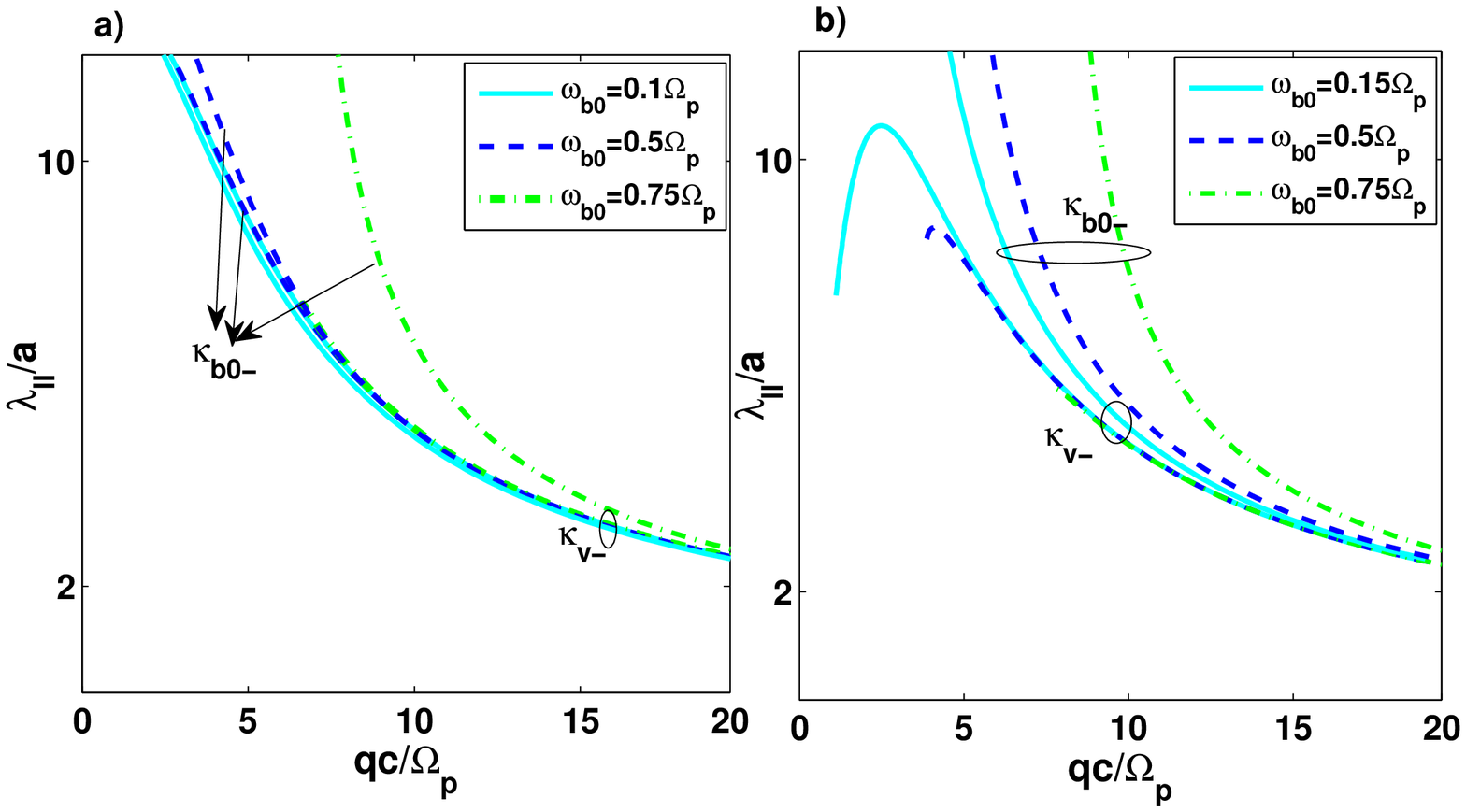}}
\caption{The normalized localization length versus SPP wave vector
for a) lower and b) higher bands of the slot waveguide in Voigt
configuration at the upper interface with $a=0.1~\mu m$, $\omega_{b}
= 0.5~\Omega_p$ and different $\omega_{b_0} = 0.1,~ 0.5,~
0.75~\Omega_p$.} \label{fig7}
\end{figure*}

To study dependence of the SPP dispersion on the parameters of the
lower WSM we have plotted them as a function of wave vector for
different $\omega_{b_0}=0.1,~0.25,~0.5,~0.75~\Omega_p$ and
$\omega_{b}=0.5~\Omega_{p}$ for $a = 0.1~\mu m$ in Fig. \ref{fig6}.
As we can see, there is no significant variation in SPP dispersion
by changing $\omega_{b_0}$, except small variations in the short
wave vectors close to the bulk plasmon dispersion curves. These
variations take place due to the change in the bulk plasmon
dispersion by changing $\omega_{b_0}$. In Figs. \ref{fig7} (a) and
(b) the normalized localization length ($\lambda_{ll}/a$) is plotted
for the lower and higher SPP bands with different $\omega_{b_0}$. As
it is apparent from these figures, in spite of the negligible
variation in the SPP modes frequency with changing $\omega_{b_0}$,
there is a significant change in the reduced decay constant of the
lower WSM. It decreases for all wave vectors by decreasing
$\omega_{b_0}$.
%
%
\begin{figure}[ht]
\centerline{\includegraphics[width=9cm]{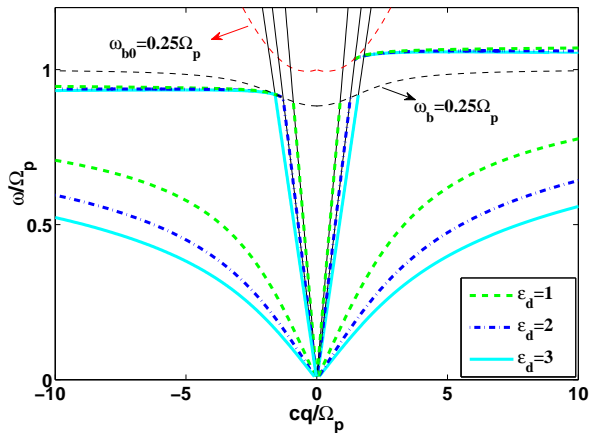}} \caption{Surface
plasmon polariton dispersion of the slot waveguide in Voigt
configuration at the upper interface for
$\omega_{b}=\omega_{b_0}=0.25~\Omega_{p}$ and for different
dielectric constants of the dielectric layer $\varepsilon_{d}=1.0,~
2.0,~ 3$.} \label{fig8}
\end{figure}
%
\begin{figure*}[ht]
\centerline{\includegraphics[width=18cm]{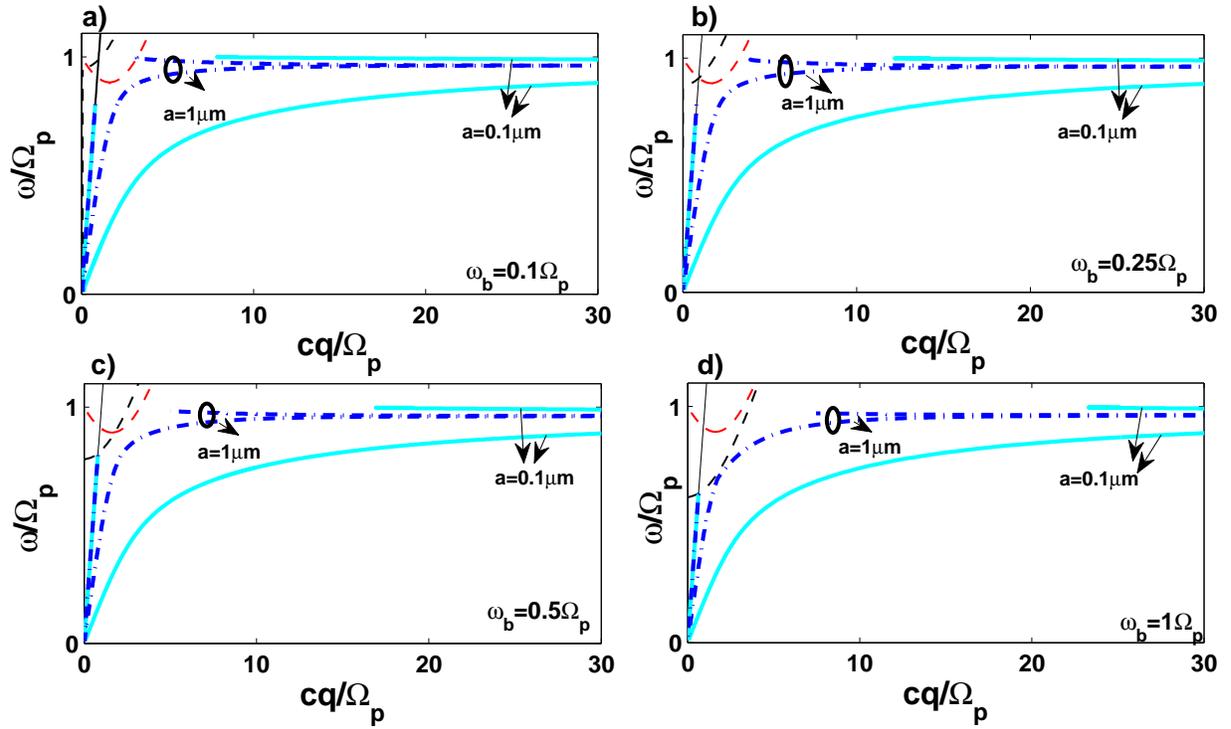}} \caption{SPP
dispersion of the slot waveguide in Faraday configuration at the
upper interface for waveguide widths $a=0.1,~ 1.0~\mu m$,
$\omega_{b_0}=0.5~\Omega_{p}$ and for different $\omega_{b}=0.1,~
0.25,~ 0.5,~ 1.0~\Omega_{p}$. The other parameters are same as Fig.
\ref{fig2} and the black and red dash lines denote the bulk plasmon
dispersions.} \label{fig9}
\end{figure*}
%
\begin{figure*}[ht!]
\centerline{\includegraphics[height=9cm,width=18cm]{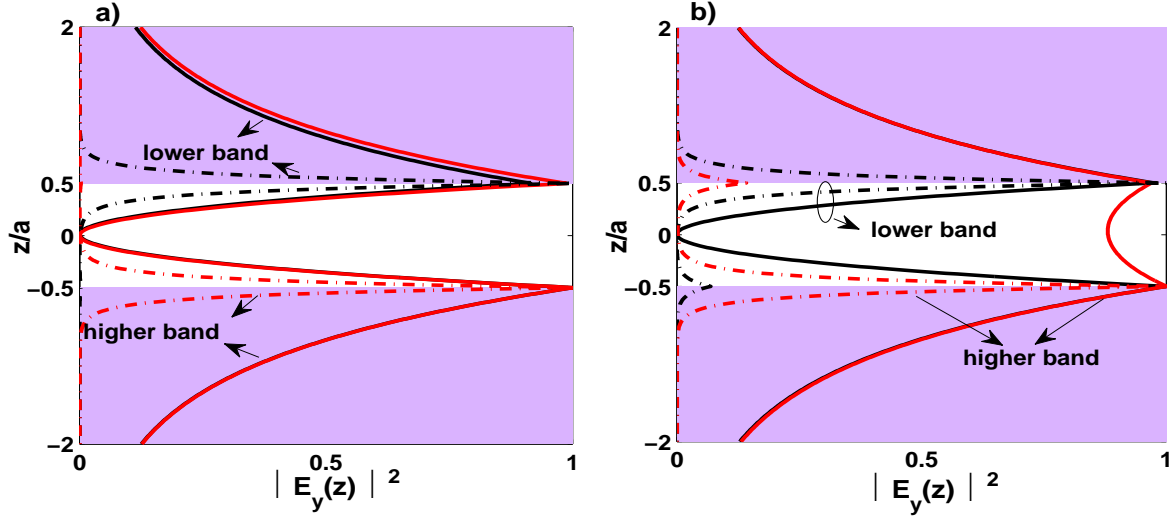}}
\caption{The normalized electric field intensities of the slot
waveguide in Faraday configuration at the upper interface by
assuming $\omega_{b_0}= 0.5~\Omega_{p}$ and for a) $\omega_{b}=
1.0~\Omega_{p}$ and b) $\omega_{b}=0.25~\Omega_{p}$ with $a=1.0~\mu
m$ and $a=0.1~\mu m$ indicated by the dash-dot and solid lines,
respectively. The red lines denote profile of the electric field for
a SPP mode of the higher band and the black ones are for the lower
band.}\label{fig10}
\end{figure*}
%
%
\begin{figure*}[ht!]
\centerline{\includegraphics[height=9cm,width=18cm]{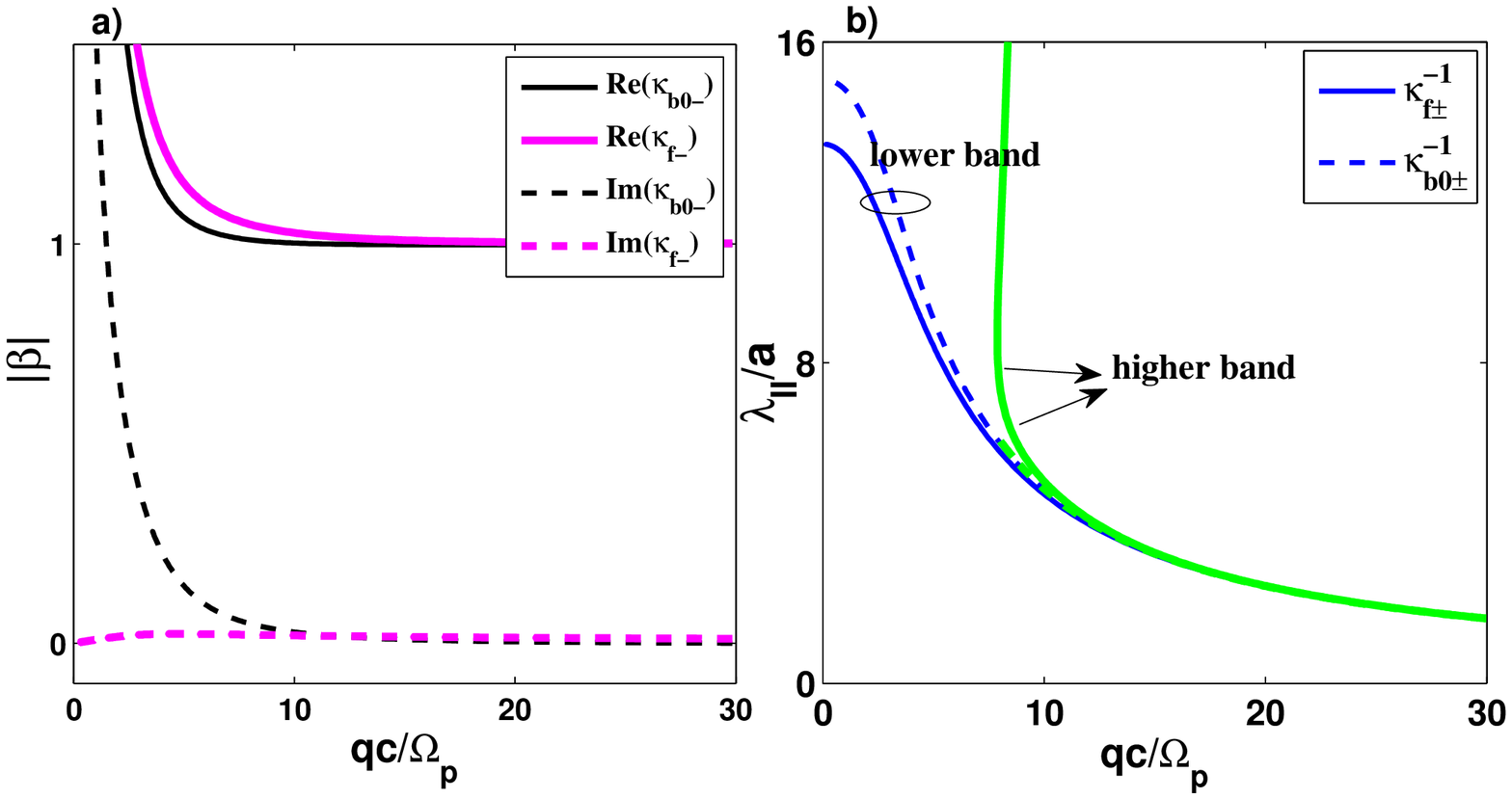}}
\caption{a) Real and imaginary parts of the reduced decay constants
versus wave vector for lower band with $\omega_{b_0}=0.5~\Omega_p$,
$\omega_b=0.1~\Omega_p$ and $a=0.1~\mu m$. b) The normalized
localization length as a function of the wave vector for both lower
and higher bands with $\omega_{b_0} = 0.5~\Omega_p$, $\omega_{b} =
0.1~\Omega_p$ and $a=0.1,~ 1.0~\mu m$.}\label{fig11}
   \end{figure*}
%

Dependence of the SPP dispersion on the dielectric constant of the
dielectric layer has been shown in Fig. \ref{fig8}, which we have
plotted SPP dispersion for $\varepsilon_{d}=1.0,~ 2.0,~ 3.0$ with
$\omega_{b}=\omega_{b_0}=0.25~\Omega_{p}$ and $a=0.1~\mu m$. By
increasing $\varepsilon_{d}$ a substantial decrease is seen in the
lower bands and the bands lying on the light line while the higher
bands remain almost intact.

{\it The slot waveguide with Faraday configuration:} Now we turn to
study SPP dispersion of a slot waveguide with Faraday configuration
at the upper interface. Since, the resultant dispersion curves are
reciprocal, thus we show dispersion curves only for $q>0$. Fig.
\ref{fig9} shows SPP dispersion for waveguide widths $a=0.1,~
1.0~\mu m$ and the other parameters same as Fig. \ref{fig2}. In this
case there are two bands both below the bulk plasmon frequency. The
lower band starts from the origin and then approaches to the
asymptotic frequency $\omega_{as}^f$ given by Eq. \ref{eq2.21}, but
the higher band is comprised of two branches with a gap between
them. The lower branch rises just to the right of the light line and
terminates when it intersects the bulk plasmon dispersion while the
higher branch is a nearly flat band which starts at the bulk plasmon
frequency and finally tends to its asymptotic frequency
$\omega_{as}^{b_0}=\Omega_p$. In Figs. \ref{fig9} (a)-(d) we have
compared dispersion curves for different waveguide widths and
different values of $\omega_{b}$. As we can see from the figures,
decreasing the waveguide width leads to shift of the lower band to
the lower frequencies and increment of the gap of the higher band.
Although, increasing $\omega_{b}$ does not alter the lower band but
it leads to enlarging the gap of the higher band.

In Figs. \ref{fig10} (a) and (b) the normalized $y$ component of the
electric field intensity has been shown for two different values of
$\omega_b=1.0,~ 0.25~\Omega_p$ at asymptotic frequencies for both
higher and lower bands with waveguide widths $a=0.1~\mu m$ and
$a=1.0~\mu m$ denoted respectively by the solid and the dash-dot
lines. As we can see, in both figures the electric filed profile is
nearly symmetric with respect to the $z$ coordinate for a small
waveguide width ($a=0.1~\mu m$ denoted by solid lines), while it is
asymmetric for a large waveguide width ($a=1.0~\mu m$ denoted by
dash-dot lines). The symmetric profile for the electric field arises
from the effective mixing of the SPP modes at two upper and lower
interfaces in small waveguide widths. However, in the limit of a
wide waveguide this hybridization is not considerable and SPP modes
belonging to lower (higher) band are highly localized at the upper
(lower) interface. The electric filed intensity is suppressed to
zero at the middle of the dielectric layer for
$\omega_b=1.0~\Omega_p$ (Fig. \ref{fig10} (a)) and for
$\omega_b=0.25~\Omega_p$ (Fig. \ref{fig10} (b)) except the SPP modes
of the higher band for small waveguide widths.

The decay constants in both upper and lower WSM media have complex
values for SPP modes of both higher and lower bands. The real parts
of them tend to $q$, but their imaginary parts approaches to zero
for large wave vectors. So, it is clear that the real parts of decay
constants are more dominant than their imaginary parts at large wave
vectors while both of them play significant rule at small wave
vectors. For example, Fig. \ref{fig11} (a) shows real and imaginary
parts of the reduced decay constants $\beta_{f-}=\kappa_{f-}/q$ and
$\beta_{b_0-}=\kappa_{b_0-}/q$ of lower band for $a=0.1~\mu m$.
Moreover, localization length in the upper and lower WSMs and for
both SPP modes of higher and lower bands decay with respect to $q$
and approaches to $2a$ at the limit of the large wave vectors. We
have shown $\lambda_{ll}/a$ in Fig. \ref{fig11} (b) as a function of
$qc/\Omega_p$ for both higher and lower bands for $a=0.1~\mu m$,
$\omega_b=0.1~\Omega_p$ and $\omega_{b_0}=0.5~\Omega_p$.

\section{Conclusion}\label{S4}
%
In summary we have investigated the dispersion, the electric field
profile and the localization length of SPP modes hosted by a WSM
waveguide constructed by two WSMs with distinct symmetries. We found
that incorporating a WSM with broken SIS drastically modifies the
SPP modes corresponding to a single interface of a WSM with broken
TRS in both Voigt and Faraday configurations. In the Voigt
configuration inclusion of the second WSM leads to a giant
unidirectional SPP modes in the frequencies above the bulk plasmon
frequency. The SPP modes supported by the waveguide structure in the
Faraday configuration are comprised of two bands including a gapfull
and one gapless band. Our investigations revealed that the SPP modes
dispersion and localization can be effectively tuned by the
waveguide characteristic parameters such as the topological
properties of two WSMs. The anomalous Hall effect and chiral
magnetic effect which depend on the separation of the Weyl nodes in
momentum space and energy determine the strength of the coupling of
the electrical and magnetical properties of WSMs. The inhomogeneous
optical response of WSMs originating from transverse or Hall
conductivity of these materials is estimated to be several orders of
magnitude larger than typical magnetic dielectrics
\cite{Kotov18,Tamaya18}. Therefore, a stable and efficient control
of SPP propagation at the interface of WSMs can be achieved by their
intrinsic topological properties. Practically, the chiral quantum
optics which deals with propagation-direction-dependent light-matter
interactions has attracted great attention recently \cite{Lodahl19}.
The chiral SPP modes reported here may be employed in developing
unidirectional optical circuits \cite{Dotsch05} and the directed
excitations in a ring lasers \cite{kravtsov99}.


\end{document}